\documentclass[ pra,showpacs,nofootinbib,draft]{revtex4}
\usepackage{dcolumn,graphicx}

\begin{document} 
 
\title{Theoretical Study of 
Pressure Broadening of Lithium Resonance Lines by Helium Atoms} 
 
\author 
{Cheng Zhu, James F. Babb and Alex Dalgarno} 
 
\affiliation
{ ITAMP, Harvard-Smithsonian Center for Astrophysics, 
60 Garden Street, Cambridge, Massachusetts 02138} 
 
\date{\today} 
 
\begin{abstract} 
 
Quantum mechanical calculations are performed of 
the emission and absorption profiles of the lithium $2s$-$2p$ 
resonance line under the 
influence of a helium perturbing gas. 
We use carefully constructed potential energy surfaces and transition 
dipole moments to compute the emission and absorption coefficients at 
temperatures from $200$ to $3000$~K at wavelengths between
$500$ nm and $1000$ nm.
Contributions from quasi-bound states are included. 
The resulting red and blue wing profiles are compared with previous 
theoretical calculations and with an experiment, carried out at
a temperature of $670$~K.
 
\end{abstract} 
 
\maketitle

%\pagestyle{empty}
%\draft
\input{epsf}
      
\section{Introduction} 
                  
It has been argued that prominent features in the spectra of 
brown dwarfs and extra-solar planets may be attributed to the
resonance lines of the alkali metal atoms, broadened by collisions with the
ambient hydrogen molecules and helium atoms
\cite{burrows01,seager00,brown01,allard03,burrows03}.
The interpretations
of the measured profiles which yield information on the temperatures,
densities, albedos and composition of the atmospheres are based on models
of the line-broadening. Recent studies of the potassium and sodium lines
have employed classical and semi-classical scattering theories 
\cite{burrows03,nefedov99,scheps75}.
Jungen and Staemmler \cite{jungen88} have obtained detailed emission
and absorption profiles (in arbitrary units) for the resonance line of
lithium for a wide range of temperatures
and Erdman et al.~\cite{erdman04} have presented measurement of the
absorption profile of lithium in a gas of helium at a temperature
of $2033$~K together with the results of a quasistatic semiclassical
theory.

In this paper, we consider the pressure broadening of the $2p$-$2s$ resonance line of
lithium arising from collisions with helium atoms. We carry out full
quantitative quantum-mechanical calculations of the emission and absorption
coefficients in the red and blue wings. We include contributions from
quasi-bound states and we allow for the variation of the transition
dipole moment with internuclear distance. We present results for
temperatures between 200K and 3000K and we compare with the emission
coefficient that has been measured at 670K \cite{scheps75} and with calculations of
Herman and Sando \cite{herman78} at 670K.

\section{Theory} 
\label{sec:method} 
 
For a system consisting of a mixture of lithium in a bath of
helium atoms, the emission spectrum of the lithium $2p$-$2s$ atomic line 
has both blue and red wings due to  
collisions with helium. 
If the gas densities are low enough that 
only binary collisions occur, 
the problem is reduced to the radiation of  
temporarily formed LiHe molecules. 
The broadened $2p$-$2s$ atomic emission line corresponds to  
transitions from the excited 
A$^2\Pi$ and B$^2\Sigma$ electronic states 
of the LiHe molecule to the ground X$^2\Sigma$ state. 
The X$^2\Sigma$ and B$^2\Sigma$ states have no bound states 
and the A$^2\Pi$ state has a shallow well which 
supports seven bound ro-vibrational levels. 
The level $v=6$ is very close to the dissociation limit and
it may not be found in practice. In any case it behaves
as if it belonged to the continuum.
We consider both  
bound-free and free-free transitions.
 
The emission coefficient is defined as the number of photons 
with frequency between $\nu$ and ($\nu+d\nu$) 
emitted per unit time per unit volume per unit frequency interval, 
which for bound-free transitions is given by 
\cite{herman78,sando71e,woerd85} 
\begin{equation} 
k_\nu^{b-f}(a\rightarrow b)=\omega n_{Li} n_{He} \frac{64 \pi^4 \nu^3}{3 h c^3} 
\frac{1}{Q_T} \frac{2 \pi}{\hbar} 
\cdot \Gamma^{bf}
\end{equation} 
and for free-free transitions by 
\cite{herman78,sando71e,woerd85} 
\begin{equation} 
k_\nu^{f-f}(a\rightarrow b)=\omega n_{Li} n_{He} \frac{64 \pi^4 \nu^3}{3 h c^3} 
\frac{1}{Q_T} \frac{2 \pi}{\hbar^3} 
\cdot \Gamma^{ff}, 
\end{equation} 
where $\omega$ is the probability that the initial excited electronic 
state of LiHe is populated by the collision,
which is $\frac{2}{3}$ for the A$^2\Pi$ state and 
$\frac{1}{3}$ for the B$^2\Sigma$ state. In Eqs.~(1) and (2), 
$n_{Li}$ and $n_{He}$ are the number densities of 
lithium and helium atoms, respectively, $c$ is the speed of light, and 
$Q_T=(2 \pi \mu k T/h^2)^{3/2}$ is 
the translational partition function for relative 
motion, $\mu$ is the reduced mass of the collision pair, 
$k$ is Boltzmann's constant and $T$ is the temperature. 
$\Gamma^{bf}$ and $\Gamma^{ff}$ are defined by 
\begin{equation} 
\Gamma^{bf}= 
\sum_v \sum_J g(v,J) 
|\langle g_b^{E_b,J}|D|g_a^{v,J}\rangle |^2, 
\end{equation} 
and  
\begin{equation} 
\Gamma^{ff}=\sum_J (2J+1) 
\int dE_a \exp(-E_a/kT)  
|\langle g_b^{E_b,J}|D|g_a^{E_a,J}\rangle |^2, 
\end{equation} 
where $g(v,J)$ is the relative population of levels $(v,J)$ of
the A$^2\Pi$ state and
$D(R)$ is the electronic transition dipole moment
which varies with the nuclear separation $R$.
The subscript $a$ denotes the initial state and $b$  
the final state. 
The functions $g^{v,J}$ and $g^{E,J}$ are, respectively, the bound 
and energy normalized free wave functions determined 
from the radial Schr\"{o}dinger equation 
\begin{equation} 
\frac{d^2g(R)}{dR^2} 
+\frac{2\mu}{\hbar^2}[E-V(R)-\frac{\hbar^2J(J+1)}{2\mu R^2}]g(R)=0. 
\end{equation} 
The bound energy $E_a^{v,J}$ and free energies $E_a$ and $E_b$ 
are measured from the dissociation limit of the excited states
and the ground state respectively.
The photon frequency $\nu$ is determined by the relation 
$h\nu=E_a-E_b+h\nu_0$, where $\nu_0$ is the 
atomic line frequency. 
In writing Eqs.~(3) and (4), we may replace the sum of the matrix
elements in which $J$ changes by $\pm 1$ by matrix elements in which
$J$ does not change.

The sum of Eqs.~(1) and (2) is the total photon emission rate.
In the limit of high densities, the bound levels of the A$^2\Pi$
state are in thermal equilibrium and $g(v,J)$ is given by
\begin{equation} 
g(v,J)=(2J+1)\exp (-E_a^{v,J}/kT).
\end{equation} 
In the limit of low densities, no bound states are populated and
only free-free transitions contribute to the emission spectrum.

For absorption, we consider free-bound and free-free
transitions from the ground X$^2\Sigma$ electronic 
state to the excited A$^2\Pi$ and B$^2\Sigma$ states. 
The absorption coefficient  
for free-bound transitions is given by 
\cite{doyle68,sando71,sando73} 
\begin{equation} 
\alpha_\nu^{f-b}(b\rightarrow a) 
=\omega n_{Li} n_{He} \frac{16 \pi^3 \nu}{3 h c} 
\frac{1}{Q_T} \frac{\pi}{\hbar}\exp[h(\nu-\nu_0)/kT] 
\cdot \Gamma^{bf}, 
\end{equation} 
and for free-free transitions by 
\cite{doyle68,sando71,sando73} 
\begin{equation} 
\alpha_\nu^{f-f}(b\rightarrow a) 
=\omega n_{Li} n_{He} \frac{16 \pi^3 \nu}{3 h c} 
\frac{1}{Q_T} \frac{\pi}{\hbar^3}\exp[h(\nu-\nu_0)/kT] 
\cdot \Gamma^{ff}. 
\end{equation} 
 
In Eqs.~(1), (2), (7) and (8), the emission and absorption 
coefficients are given in terms of the number of photons
in a frequency interval $d\nu$.  
The photon energies emitted or absorbed are obtained by
multiplying $k_\nu$ or $\alpha_\nu$ by $h\nu$. The 
coefficients per unit wavelength are obtained by multiplying
by $\nu^2/c$.

\section{Potentials and dipole moments} 
 
\begin{figure} 
\centerline{\epsfxsize=12.0cm  \epsfbox{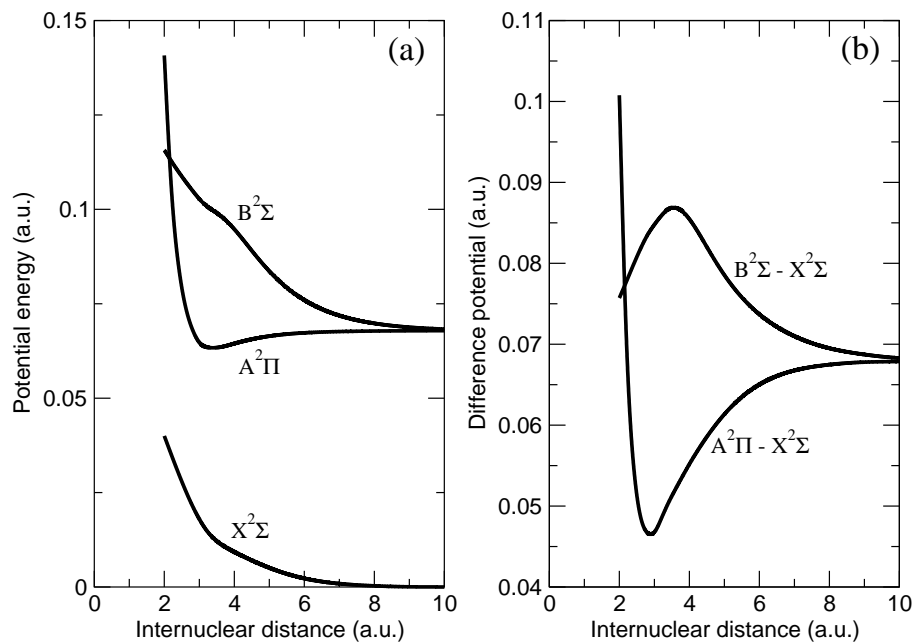}} 
\caption{ 
(a) Adopted potentials $V(R)$ in atomic units
for the $X^2\Sigma$, $A^2\Pi$ and 
$B^2\Sigma$ states. (b) Difference potentials $V_{A^2\Pi}(R)-V_{X^2\Sigma}(R)$ 
and $V_{B^2\Sigma}(R)-V_{X^2\Sigma}(R)$ in atomic units. 
} 
\label{pot} 
\end{figure}

The potential energy surfaces were constructed to 
reflect recent calculations.
For short and intermediate 
internuclear distances $R$, the energy points were taken from 
Ref. \cite{czuchaj95} for $3.0\le R \le 8.0$, 
Ref. \cite{staemmler97} for $9.0\le R \le19.0$ and 
Ref. \cite{jeung00} for $R=20.0$ 
for the $X^2\Sigma$ state, from 
Ref. \cite{jeung00} for $2.0\le R\le 2.75$ and 
Ref. \cite{behmenburg96} for $3.0\le R\le 20.0$ 
for the $A^2\Pi$ state, and from 
Ref. \cite{krauss71} for $R=2.0$ and 
Ref. \cite{jeung00} for $3.0\le R\le 20.0$ 
for the $B^2\Sigma$ state. 
At large $R$, the potential energies vary as $-C_6/R^6$. 
The theoretical values of $C_6$ of 
$22.51$ for the $X^2\Sigma$ state~\cite{yan96}, 
$28.27$ for the $A^2\Pi$ state~\cite{yan01} and 
$50.69$ for the $B^2\Sigma$ state~\cite{yan01} were adopted. 
For each potential energy curve, the data in different segments were smoothly connected 
with a cubic spline procedure. The potential energy curves 
are shown in Fig.~\ref{pot}, along with the 
difference potentials $V_{A^2\Pi}(R)-V_{X^2\Sigma}(R)$ 
and $V_{B^2\Sigma}(R)-V_{X^2\Sigma}(R)$.
In Fig.~\ref{wave},
we present the energy separations as wavelengths $\lambda$ obtained
with the classical assumption that the emission occurs through
vertical transitions as the particles move along the potential energy
curves.
The vibrational energy levels of the A$^2\Pi$ state are listed
in Table \ref{vibr}.
Values of the difference of energy of the $v=0$ and $1$ vibrational
levels of the A$^2\Pi$ state have been derived from spectroscopic
experiments~\cite{lee91}. The measured values are $289.9(4)$ cm$^{-1}$
and $290.0(3)$ cm$^{-1}$, obtained by differencing two
A$^2\Pi$--D$^2\Delta$ transitions.
From Table \ref{vibr}, we obtain $288.4$
cm$^{-1}$.
An earlier calculation by Jungen and Staemmler~\cite{jungen88}
yielded $260$ cm$^{-1}$.

\begin{figure} 
\centerline{\epsfxsize=12.0cm  \epsfbox{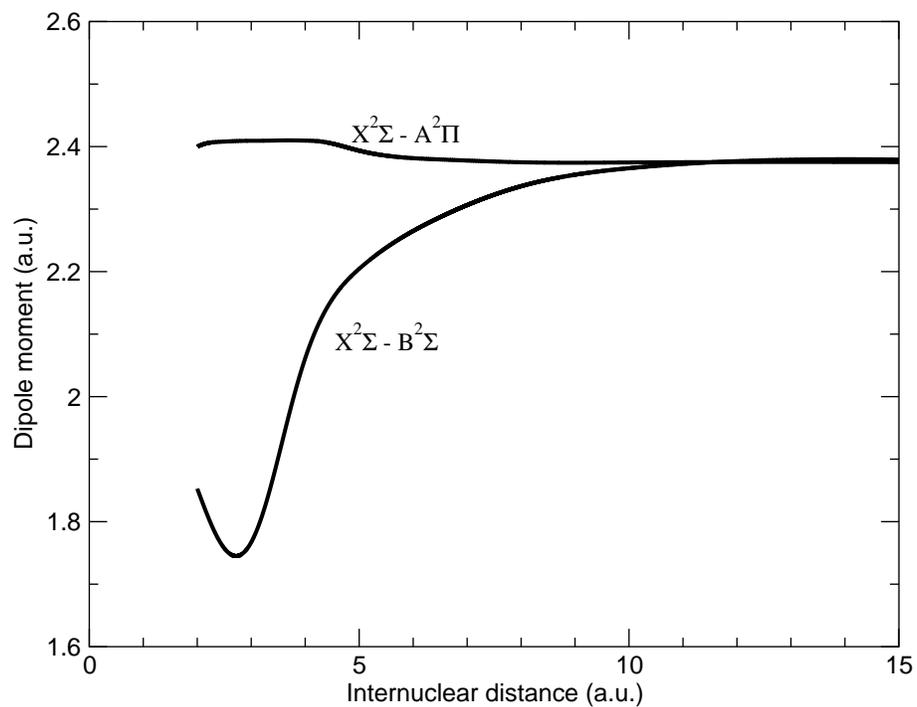}} 
\caption{ 
Adopted dipole moment curves for the ${X^2\Sigma}-{A^2\Pi}$ 
and ${X^2\Sigma}-{B^2\Sigma}$ transitions. 
} 
\label{dip} 
\end{figure}

The adopted dipole moment curves~\cite{grycuk}  
are shown in Fig.~\ref{dip}.

\section{Calculations} 

The radial equation (5) was solved using the Numerov method.
The integrands of the free-free matrix elements oscillate at large
internuclear distances and the integrals fail to converge. They can be
transformed by the equation
\begin{equation}
D(R)=\{ D(R)-D(\infty) \} + D(\infty).
\end{equation} 
The matrix element of $\{ D(R)-D(\infty) \}$ converges because
the operator tends to zero at large $R$. The matrix element
of the constant $D(\infty)$ may be written in the form\cite{herman78}
\begin{equation} 
D(\infty)\langle g_b^{E_b,J}|g_a^{E_a,J}\rangle =
D(\infty)\langle g_b^{E_b,J}|\Delta V|g_a^{E_a,J}\rangle/(E_a-E_b), 
\end{equation} 
where $\Delta V=V_a(R)-V_b(R)$ is the difference potential.
The right hand side of (9) is convergent since  
$\Delta V\rightarrow 0$ for $R\rightarrow \infty$.

The integration over energy in Eq.~(4) was carried out using
the Gauss-Laguerre method with $100$ points.
We also ran the code with $200$ points and found no difference in the
results. We included values of $J$ up to $50$ for $T=200$~K and up to
$200$ for $T=3000$~K.

Shape resonances occur in scattering by the A$^2\Pi$ potential. They
were determined by calculations at high energy resolution. Table \ref{qlevels} is
a list of the values of $J$ and the locations and widths $\gamma_r^{v,J}$
of the quasi-bound levels. We used Simpson's rule with energies
selected at closely spaced intervals to evaluate the resonance
contributions to the energy integrals.

\section{Results and discussions} 
 
\begin{figure} 
\centerline{\epsfxsize=14.0cm  \epsfbox{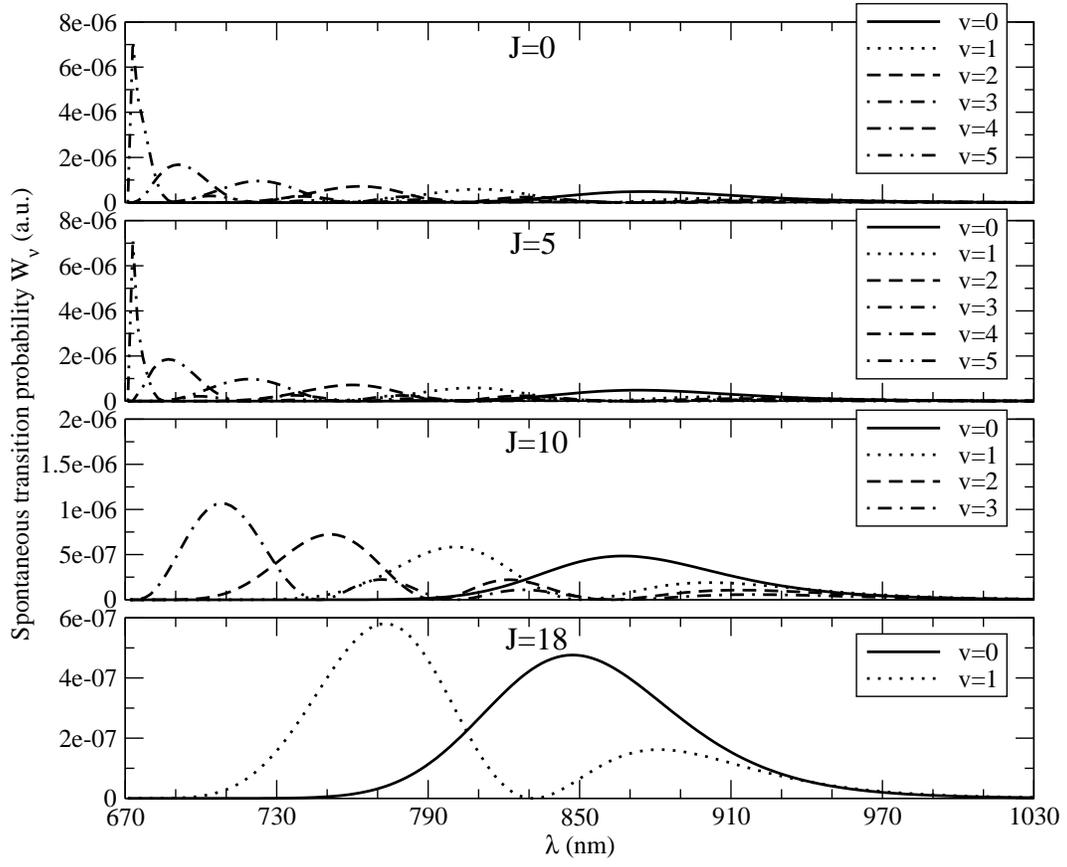}} 
\caption{Spontaneous transition probabilities
$W_\nu$, Eq.~(11), in atomic units for vibrational levels 
$v=0-5$ for $J=0$ (upper panel), $5$ (upper middle panel),
$10$ (lower middle panel) and $18$ (lower panel). 
} 
\label{omega_nv} 
\end{figure}

The bound-free emission coefficient (1) is a weighted sum of the
emission rates of the individual ro-vibrational levels of 
the A$^2\Pi$ state. The spontaneous transition probabilities are given by 
\begin{equation} 
W_\nu= \frac{64 \pi^4 \nu^3}{3 h c^3} 
\frac{2 \pi}{\hbar} 
|\langle g_b^{E_b,J}|D|g_a^{v,J}\rangle |^2. 
\end{equation}
In Fig.~\ref{omega_nv} we plot 
$W_\nu$ in atomic units 
as a function of wavelength for rotational states $J=0$, $5$, $10$ and $18$
of the vibrational levels $v=0$ to $v=5$. The corresponding  lifetimes in seconds
are listed in Table \ref{life}.
For high vibrational levels, they are approaching the limit of the
radiative lifetime of the resonance state of the lithium atom which is
$27.2$~ns~\cite{marinescu95}.

\begin{figure} 
\centerline{\epsfxsize=12.0cm  \epsfbox{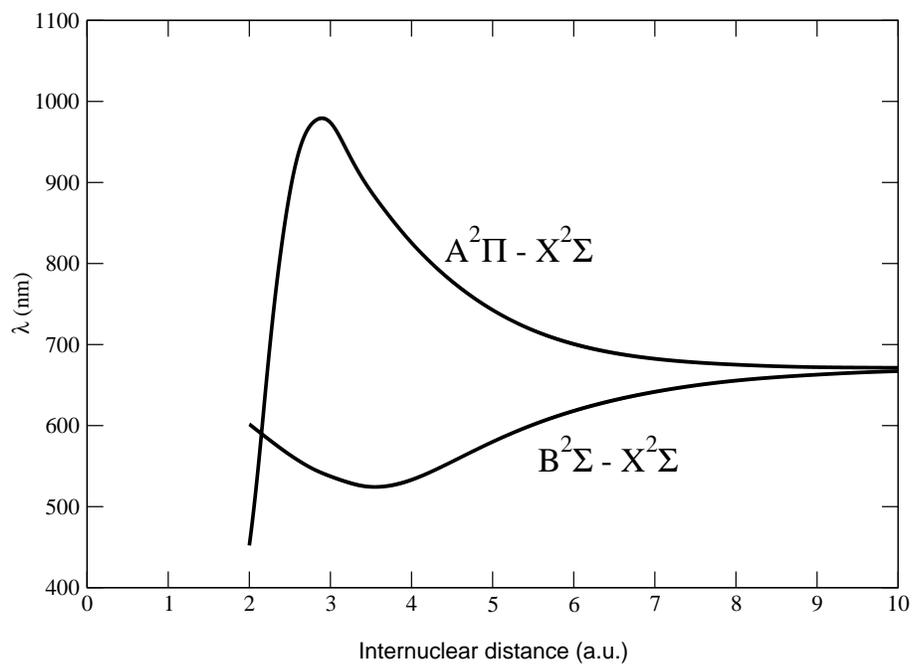}} 
\caption{Wavelengths converted from difference potentials
for A$^2\Pi$-X$^2\Sigma$ and B$^2\Sigma$-X$^2\Sigma$
transitions.
} 
\label{wave} 
\end{figure}

\begin{figure} 
\centerline{\epsfxsize=12.0cm  \epsfbox{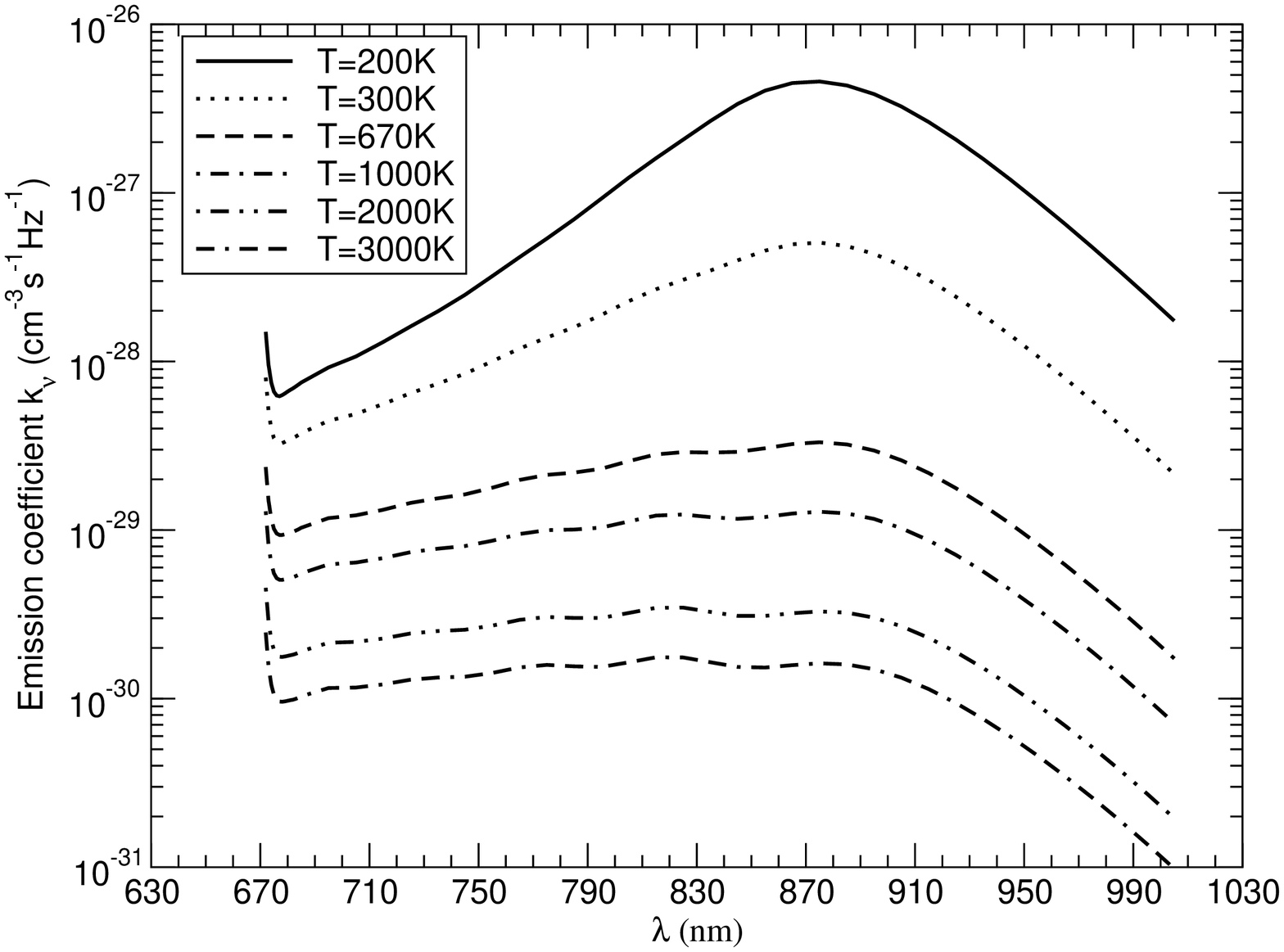}} 
\caption{Contributions of bound-free transitions 
to the total emission coefficients at temperatures 
$T=200$, $300$, $670$, $1000$, $2000$ 
and $3000$~K. 
Unit gas densities are used, 
$n_{Li}=n_{He}=1$~cm$^{-3}$. 
} 
\label{emi_bound} 
\end{figure}

\begin{figure} 
\centerline{\epsfxsize=14.0cm  \epsfbox{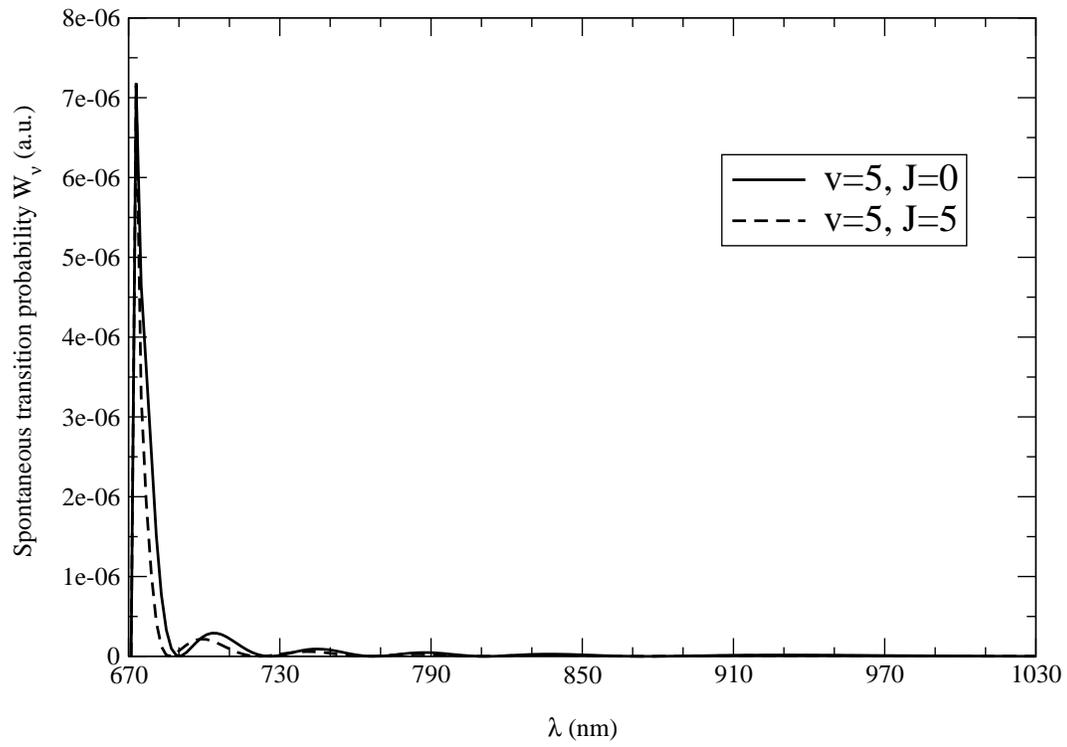}} 
\caption{Spontaneous transition probabilities
$W_\nu$, Eq.~(11), in atomic units for ro-vibrational levels 
$(v,J)=(5,0)$ and $(5,5)$. 
} 
\label{omega_nv2} 
\end{figure}

In Fig.~\ref{emi_bound} we present the contributions to the emission
coefficient $k_\nu$ in units of cm$^{-3}$s$^{-1}$Hz$^{-1}$ of the A$^2\Pi$ state
from the bound-free
transitions, assuming that the levels are populated in thermal
equilibrium at temperatures ranging from $200$~K to $3000$~K.
The contributions from the resonances were multiplied by 
$\gamma_r^{v,J}/(\gamma_r^{v,J}+\gamma_i^{v,J})$, 
where $\gamma_r^{v,J}$ and $\gamma_i^{v,J}$
are the resonance and radiative decay widths\cite{sando71e},
but the effect is negligible.
At no temperature and at no frequency is the quasi-bound contribution
more than a few percent of the total.
The shape of $k_\nu$ results from a superposition of the spectra
in Fig.~\ref{omega_nv} weighted by the populations of the
ro-vibrational levels. The sharp increase near the resonance
wavelength at $670.8$ nm is the contribution from low-lying
rotational states of vibrational level $v=5$, illustrated in
Fig.~\ref{omega_nv2}.
A peak in the spectrum at around $870$ nm is apparent at low
temperatures but it tends to smooth out at high temperatures.
At the higher temperatures, there is a plateau region associated
with emission from the attractive region of the potential.
The emission decreases rapidly at long wavelengths because there
are no values of the internuclear distance $R$ that correspond to
wavelengths beyond $1000$ nm, as shown by Fig.~\ref{wave}.

\begin{figure} 
\centerline{\epsfxsize=12.0cm  \epsfbox{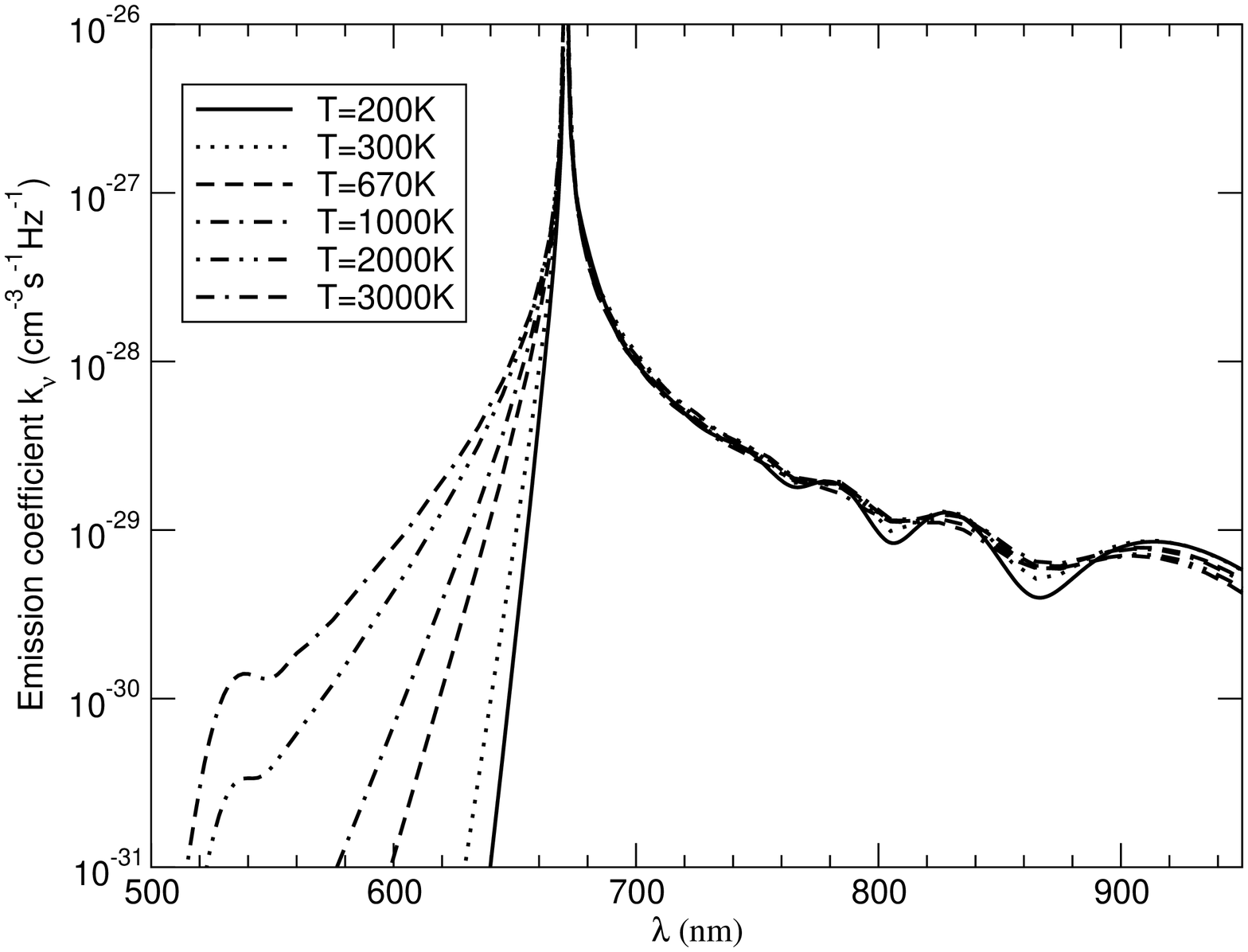}} 
\caption{Contributions of free-free transitions 
to the total emission coefficients at temperatures 
$T=200$, $300$, $670$, $1000$, $2000$ 
and $3000$~K. 
Unit gas densities are used, 
$n_{Li}=n_{He}=1$~cm$^{-3}$. 
} 
\label{emi_free} 
\end{figure}

The calculated free-free photon emission rates $k_\nu$ in units of
cm$^{-3}$s$^{-1}$Hz$^{-1}$ produced in transitions from the A$^2\Pi$
state to the X$^2\Sigma$ state are presented in Fig.~\ref{emi_free}. Weak
quantum oscillations are found that correspond to transitions from
the attractive region of the A$^2\Pi$ potential. There occur rapid
increases in $k_\nu$ at wavelengths close to the atomic resonance
wavelength. At some point as we approach the line center the binary
approximation that we have adopted breaks down. Our calculations are
valid only in the wings of the line.
Studies of the profile at line center have been reviewed by
Lewis~\cite{lewis80}.

The blue wings arise from the B$^2\Sigma$-X$^2\Sigma$ transitions. Only
free-free transitions occur and the result is a smooth profile
decreasing rapidly with decreasing wavelength except for a weak
satellite feature near $536$ nm that is apparent at high
temperatures. The behavior is expected from a consideration
of Fig.~\ref{wave} which shows that internuclear distances in the
B$^2\Sigma$ state that correspond to short wavelengths are
inaccessible.
The satellite feature is produced by emission from internuclear distances
at which the two potential energy curves are parallel~\cite{burrows03}.
Previous work has located it at $516$ nm \cite{krauss71} or
$506$ nm \cite{scheps75}. There is an indication of the satellite near
$530$ nm in an experimental spectrum obtained by Lalos
and Hammond~\cite{lalos62}.

\begin{figure} 
\centerline{\epsfxsize=12.0cm  \epsfbox{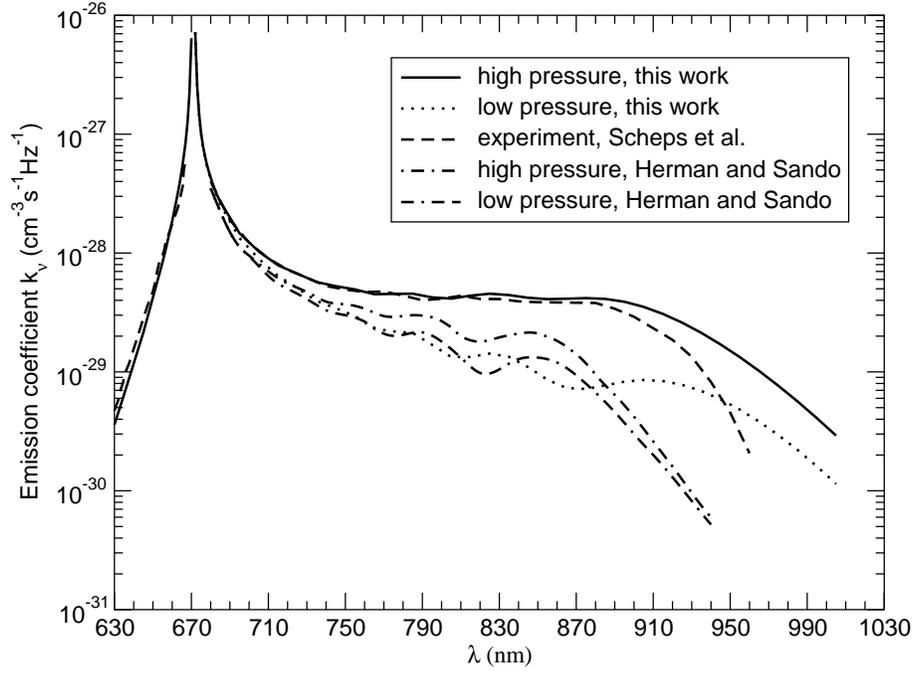}} 
\caption{ 
Comparisons of our calculated emission coefficients
with experimental measurement of Scheps et al. 
\protect\cite{scheps75} 
and previous theoretical calculations of Herman and Sando
\protect\cite{herman78}. 
For the blue wing, the high and low pressure curves are identical.
Unit gas densities are used, 
$n_{Li}=n_{He}=1$~cm$^{-3}$. 
} 
\label{withHerman} 
\end{figure}

Semi-classical calculations by Jungen and Staemmler~\cite{jungen88}
over a range of temperatures above $200$ K have similar profile
shapes to those we obtain with quantum-mechanical calculations.
Measurements of the line profile at $670$ K at high pressures
have been carried out by Scheps et al.~\cite{scheps75}
and calculations at high and low pressures in the red wing
have been made by Herman and Sando~\cite{herman78}.
In Fig.~\ref{withHerman} we give a comparison with our results. There is
broad agreement between the theoretical low pressure profiles
which arise from the free-free transitions.
The differences in detail can be ascribed to the adopted
interaction potentials. The theoretical calculations of the
high pressure profiles have the same shape. They both have a
plateau at intermediate wavelengths, increase sharply near the
line center and fall off rapidly far in the red wings.
They differ quantitatively in the plateau region.
Agreement between the present theoretical red and blue wings
and experiment is close and it indicates that the binary
approximation is valid to within $20$ nm of line center.
There may be a discrepancy at wavelengths longer than $950$ nm
where the emission coefficient is becoming very small.
According to Scheps et al.~\cite{scheps75} their data at
long wavelengths are noisy. The long range interactions should
be reliable~\cite{staemmler97} and we have
explored the effects of modifying the X$^2\Sigma$ potential
in the short range repulsive region. However we were unable to obtain
precise agreement with the measurements.

\begin{figure} 
\centerline{\epsfxsize=12.0cm  \epsfbox{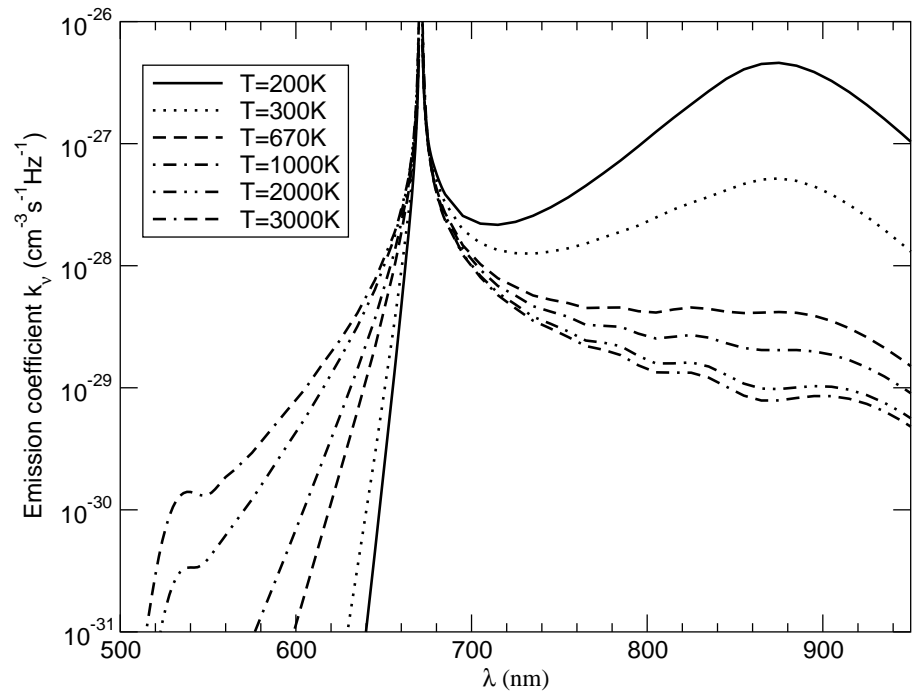}} 
\caption{High pressure emission coefficients at temperatures 
$T=200$, $300$, $670$, $1000$, $2000$ 
and $3000$~K. 
Unit gas densities are used, 
$n_{Li}=n_{He}=1$~cm$^{-3}$. 
} 
\label{emi_sum} 
\end{figure}

In Fig.~\ref{emi_sum}, we present the red and blue wings of the line
profiles in the high pressure limit for temperatures between $200$~K and $3000$~K.
The total energy emitted per second in the red wing
at wavelengths longer than $690.8$~nm, given by
\begin{equation}
I_1=\int_0^{\nu_0-\Delta_1}h\nu k_\nu d\nu,
\end{equation}
where $\Delta_1$ corresponds to a wavelength
shift of $20$ nm red from the line center,
and the total energy emitted per second in the blue
wing at wavelengths shorter than $650.8$ nm, given by
\begin{equation}
I_2=\int_{\nu_0+\Delta_2}^\infty h\nu k_\nu d\nu,
\end{equation}
where $\Delta_2$ corresponds to a wavelength
shift of $20$ nm blue from the line center,
are listed in Table \ref{inten}
for temperatures between $200$~K and $3000$~K.

\begin{figure} 
\centerline{\epsfxsize=12.0cm  \epsfbox{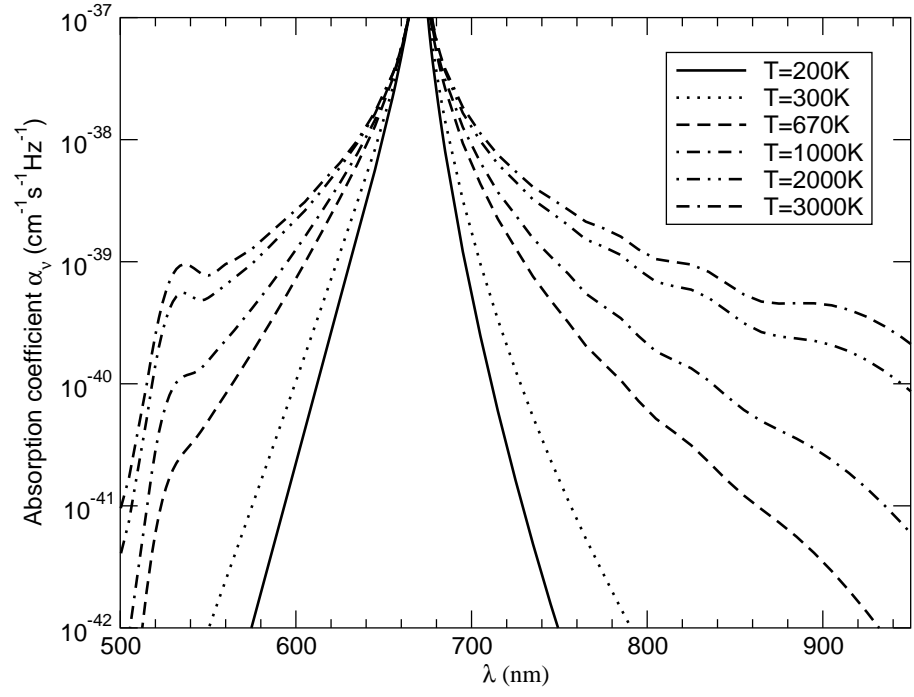}} 
\caption{ 
Absorption coefficients at temperatures 
$T=200$, $300$, $670$, $1000$, $2000$ 
and $3000$~K. 
Unit gas densities are used, 
$n_{Li}=n_{He}=1$~cm$^{-3}$. 
} 
\label{abs_sum} 
\end{figure}

The absorption coefficients
are shown
in Fig.~\ref{abs_sum}. They are similar in shape to the semi-classical
calculations of Jungen and Staemmler~\cite{jungen88}
and of Erdman et al.~\cite{erdman04}.
There is a satellite near $536$ nm which also appears in the emission
spectrum but otherwise the absorption coefficients decrease steadily
with separation from the position of the resonance line.
There is an indication of a satellite near $600$~nm
in the calculation of Erdman et al.~\cite{erdman04}
based on different potential curves which may
have the same origin.
 
\section{Conclusions} 
 
We have carried out quantum mechanical calculations for 
the Li-He emission and absorption spectra at 
temperatures from $200$~K to $3000$~K
and wavelengths from
$500$ nm to $1000$ nm.
We find a blue satellite near $536$~nm 
for $T=3000$~K and  a red satellite near $870$~nm 
for $T\approx 200$--$300$~K
in the emission spectra,
and a blue satellite  near $536$~nm
for $T\approx 2000$--$3000$~K in the absorption spectra.
At $670$~K, our emission coefficients are in good agreement with
experiment. 
 
\acknowledgments 
 
This work was supported in part by 
the NSF through 
a grant for ITAMP to the Smithsonian 
Astrophysical Observatory 
and Harvard University and by NASA under award NAG5-12751. 
We are grateful to Dr. T. Grycuk and to Dr. G.-H. Jeung
for sending us details of the transition dipole moments.

\newpage 
\begin{table}[h]
\caption{Vibrational energy levels of the A$^2\Pi$ state of the
LiHe molecule.
}
\label{vibr}
\end{table} 
\begin{center} 
\begin{tabular}{cccccccc} \hline\hline
$v$                & ~~~~0~~~~ & ~~~~1~~~~ & ~~~~2~~~~ & ~~~~3~~~~
& ~~~~4~~~~ & ~~~~5~~~~ & ~~~~6~~~~  \\
\hline
$E^{v,J=0}$ (cm$^{-1}$) & $-837.7$  & $-549.3$  & $-327.7$
& $-168.2$  & $-67.11$  & $-15.69$  & $-0.9170$   \\
\hline\hline 
\end{tabular} 
\end{center}

\newpage 
\begin{table}[h]
\caption{Resonance energies $E_r^{v,J}$ and widths $\gamma_r^{v,J}$
of the quasi-bound levels of the A$^2\Pi$ state.
}
\label{qlevels}
\end{table} 
\begin{center} 
\begin{tabular}{cccc} \hline\hline
~~~~~$J$~~~~~ &~~~~~$v$~~~~~& ~~~~~$E_r^{v,J}$ (Hartree)~~~~~
& ~~~~~$\gamma_r^{v,J}$ (Hartree)~~~~~ \\
\hline
  2&1&      2.14$\times$10$^{-7}$ & 2.1$\times$10$^{-8}$\\
  6&1&      1.34$\times$10$^{-5}$ & 6.5$\times$10$^{-7}$\\
  9&1&      1.04$\times$10$^{-5}$ & 7.1$\times$10$^{-13}$\\
 10&1&      6.56$\times$10$^{-5}$ & 1.2$\times$10$^{-6}$\\
 11&1&      1.19$\times$10$^{-4}$ & 1.8$\times$10$^{-5}$\\
 13&1&      9.15$\times$10$^{-5}$ & 2.3$\times$10$^{-9}$\\
 14&1&      1.94$\times$10$^{-4}$ & 2.1$\times$10$^{-6}$\\
 15&1&      2.91$\times$10$^{-4}$ & 2.4$\times$10$^{-5}$\\
 16&1&      1.23$\times$10$^{-4}$ & 1.3$\times$10$^{-12}$\\
 17&1&      2.90$\times$10$^{-4}$ & 7.0$\times$10$^{-8}$\\
 18&1&      4.50$\times$10$^{-4}$ & 6.0$\times$10$^{-6}$\\
 19&1&      1.82$\times$10$^{-4}$ & 3.6$\times$10$^{-15}$\\
 20&1&      4.21$\times$10$^{-4}$ & 1.6$\times$10$^{-9}$\\
 21&1&      6.57$\times$10$^{-4}$ & 7.0$\times$10$^{-7}$\\
 22&1&      2.84$\times$10$^{-4}$ & 6.2$\times$10$^{-17}$\\
 22&2&      8.81$\times$10$^{-4}$ & 1.7$\times$10$^{-5}$\\
 23&1&      6.07$\times$10$^{-4}$ & 3.4$\times$10$^{-11}$\\
 24&1&      9.31$\times$10$^{-4}$ & 4.2$\times$10$^{-8}$\\
 25&1&      1.25$\times$10$^{-3}$ & 3.0$\times$10$^{-6}$\\
 26&1&      1.55$\times$10$^{-3}$ & 3.1$\times$10$^{-5}$\\
\hline\hline 
\end{tabular} 
\end{center}

\newpage 
\begin{table}[h]
\caption{Lifetimes (ns) of selected ro-vibrational levels $(v,J)$ of
the A$^2\Pi$ state.
}
\label{life}
\end{table} 
\begin{center} 
\begin{tabular}{ccccccc} \hline\hline
 & ~~~~$v=0$~~~~ & ~~~~$v=1$~~~~ & ~~~~$v=2$~~~~ & ~~~~$v=3$~~~~
& ~~~~$v=4$~~~~ & ~~~~$v=5$~~~~  \\
\hline
$J=0$  & 59.2  & 51.5  & 44.1 & 37.7  & 32.4  & 28.6     \\
$J=5$  & 58.8  & 50.9  & 43.5 & 37.0  & 31.7  & 27.7     \\
$J=10$ & 57.4  & 49.4  & 41.8 & 35.0  &   &      \\
$J=18$ & 53.0  & 43.7  &  &   &   &      \\
\hline\hline 
\end{tabular} 
\end{center}

\newpage 
\begin{table}[h]
\caption{Total energy emitted per second (ergs cm$^{-3}$ s$^{-1}$)
defined in Eqs.~(12) and (13) for the red and blue wings, respectively.
}
\label{inten}
\end{table} 
\begin{center} 
\begin{tabular}{ccc} \hline\hline
   T(K)  &  ~~~~~Red wing~~~~~   & ~~~~~Blue wing~~~~~    \\
\hline
  200 &    4.68$\times$10$^{-25}$ &     1.46$\times$10$^{-29}$\\
  300 &	   7.59$\times$10$^{-26}$ &     9.79$\times$10$^{-29}$\\
  670 &	   1.64$\times$10$^{-26}$ &     8.27$\times$10$^{-28}$\\
 1000 &	   1.22$\times$10$^{-26}$ &     1.59$\times$10$^{-27}$\\
 2000 &	   9.39$\times$10$^{-27}$ &     3.51$\times$10$^{-27}$\\
 3000 &	   8.67$\times$10$^{-27}$ &     4.88$\times$10$^{-27}$\\
\hline\hline 
\end{tabular} 
\end{center}

\clearpage
\newpage

\clearpage
\newpage

\clearpage
\newpage 
 
\clearpage
\newpage 
 
\clearpage
\newpage 
 
\clearpage
\newpage 
 
\clearpage
\newpage 
 
\clearpage
\newpage 
 
\clearpage
\newpage 
 
\clearpage
\newpage 
 
\end{document}